\documentclass[notitlepage]{article}
\usepackage{graphicx}
\usepackage{booktabs}
\usepackage{subfig}
\usepackage{array}
\usepackage{upgreek}
\usepackage{amsmath}
\usepackage{fullpage}
\begin{document}
\title{Response to comment on theoretical RF field limits of multilayer coating structures of superconducting resonator cavities}

\author{Sam Posen\thanks{sep93@cornell.edu}, Matthias U. Liepe$^1$, Gianluigi Catelani$^2$, James P.~Sethna$^3$,\\
and Mark K. Transtrum$^4$\\
$^1$Cornell Laboratory for Accelerator-Based Sciences and Education, Ithaca, NY, USA\\
$^2$Forschungszentrum J{\"u}lich Peter Gr{\"u}nberg Institut (PGI-2), J{\"u}lich, Germany\\
$^3$LASSP, Physics Department, Clark Hall, Cornell University, Ithaca, NY, USA\\
$^4$ Department of Bioinformatics and Computational Biology,\\
University of Texas M. D. Anderson Cancer Center, Houston, Texas, USA}

\maketitle

\noindent\emph{Abstract}

A comment to the authors' SRF Conference pre-print \cite{orig} was submitted by A. Gurevich to the arXiv \cite{gcomment}. In this response, we show that the arguments used in the comment are not valid.

\section{Summary Points}
In A. Gurevich's comment \cite{gcomment}, he presents three points aiming to summarize the conclusions of \cite{orig}. All three summarizing points are incorrect. The incorrect points are repeated and the misinterpretations are presented below.

\begin{enumerate}
\item \emph{``The lower critical field Hc1 of SIS multilayers is zero so they do not protect the SRF cavities against magnetic flux penetration by enhancing Hc1 in thin layers, which according to Ref. 1, was the main point of Ref. 2.''} [Ref 2 refers to \cite{APL}.]

The authors never claimed that SIS multilayers do not protect SRF cavities against magnetic flux penetration. In fact, it was shown that the superheating field of the SIS structure can, for optimal thicknesses, be slightly higher than the bulk value of the film material.

\item \emph{``SIS multilayers do not significantly enhance the superheating field $H_{sh}$ as compared to Nb even if the S layers are made of materials (like Nb$_3$Sn) with the thermodynamic critical field Hc much higher than $H_c$ of Nb.''}

The authors clearly show that the superheating field of SIS multilayers could be significantly higher than the superheating field of niobium. This can be seen in their Figure 3. The point the authors were making is that the superheating field of SIS multilayers is not significantly higher than the bulk superheating field of the film material (like Nb$_3$Sn). This summary point greatly mischaracterizes the work presented in \cite{orig}.

\item \emph{``Thin film SIS multilayers exhibit extremely high rf dissipation due to penetration of vortices. According to Ref. 1, dissipation in SIS multilayers is much stronger than in a thick Nb3Sn film deposited on the inner surface of a Nb cavity.''}

\cite{orig} does not say that the dissipation is stronger for SIS multilayers than for a thick film. Rather, a calculation is performed showing that the resulting dissipation would be unmanageable for SRF applications. This dissipation is never compared to vortex dissipation in thick films.
\end{enumerate}

\section{Discussion}
Gurevich argues against each of the incorrect summary points individually. In this section, we try to address those arguments, and explain the origins of the misinterpretations.

\subsection{$H_{c1}$ in thin film SIS multilayers}
Gurevich points out that once a vortex has passed through the superconducting film, trapping flux in the insulating layer, it is no longer dissipative, that there are no ``normal cores oscillating under the rf field.'' However, in \cite{orig}, the dissipative mechanism of concern is the drag experienced by the vortices as they pass through the film every half cycle. As shown in \cite{orig}, the dissipation due to this mechanism is unmanageable for SRF applications. In other words, Gurevich criticizes thick films for operating in a ``highly metastable state at $H>H_{c1}$ being protected by only the Bean-Livingston surface barrier''; however the SIS structure is also metastable to strongly dissipative vortex penetration.

\subsection{Superheating field in SIS multilayers}
\begin{figure}[htbp]
\begin{center}
\includegraphics[width=0.96\textwidth,angle=0]{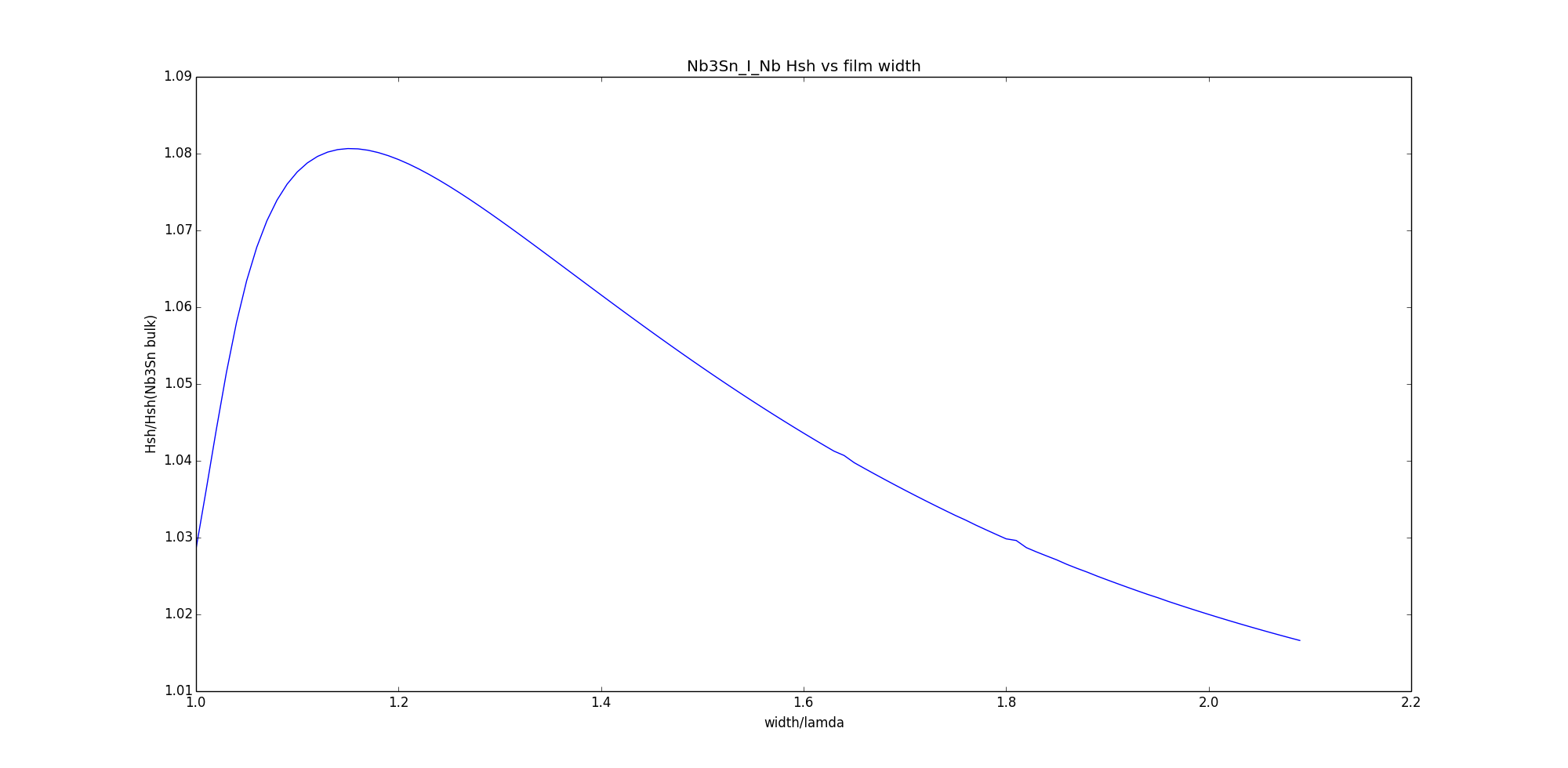}
\end{center}
\caption{Full Ginzburg-Landau calculation for a Nb$_3$Sn on Nb SIS structure agrees well with the simple London calculation.} 
\label{MarkCalc}
\end{figure}

In this section and the one previous, Gurevich criticizes the use of the London model to calculate the superheating field and recommends the use more sophisticated tools such as Ginzburg Landau theory. The authors had been working on a Ginzburg Landau calculation, and a preliminary result is shown in Figure \ref{MarkCalc}.  It shows very similar conclusions to the London model: a $\sim$10\% increase in the superheating field of the SIS structure if the film thicknesses are in the correct range. This indicates that the London theory can at least approximate the physics of the situation, in order to compare the SIS structure to thick films.

Also for this point in \cite{gcomment}, Gurevich demonstrates that the SIS structure can reach superheating fields significantly higher than that of niobium, but as was mentioned above, \cite{orig} never argued against this being true.

\subsection{Vortex dissipation in SIS multilayers}
Gurevich suggests a factor of $d/\lambda$ change in dissipation in a SIS structure compared to a thick film, but the factor would have to be very small in order to make the dissipation manageable for SRF applications. As \cite{orig} shows, if this factor is made very small, the superheating field would decrease dramatically.

Also in this section, Gurevich quotes the phrase ``unimaginably high,'' and attributes it to \cite{orig}. However, this phrase was never used. Perhaps the word ``unmanageable'' was misread.

\subsection{Comparison of thick films with SIS multilayers}
In the last point in \cite{gcomment}, it is suggested that small thermal conductivities will make the SIS structure preferable to thick films. This point was illustrated with a calculation for a Nb$_3$Sn layer many tens of penetration depths thick, using the case of a film fabricated via vapor deposition as an example. However, if high enough fields were reached that thermal runaway became a problem, a much thinner film could be used directly on top of a niobium substrate---this would have a much smaller thermal resistance. The argument in \cite{gcomment} does not present a compelling reason for including an insulating layer between the Nb$_3$Sn and the Nb, or for separating the Nb$_3$Sn into multi-layers separated by insulators.

\section{Conclusions}
The arguments invoked by the comment \cite{gcomment} to \cite{orig} show significant misinterpretation of the arguments used by the authors. The original conclusions remain valid.


\begin{thebibliography}{9}

\bibitem{orig}
Sam Posen, Gianluigi Catelani, Matthias U. Liepe, James P. Sethna, and Mark K. Transtrum, ``Theoretical field limits for multi-layer superconductors,'' arXiv:1309.3239 [physics.acc-ph] (2013).

\bibitem{gcomment}
Alex Gurevich, ``On the theoretical RF field limits of multilayer coating structures of superconducting resonator cavities,'' arXiv:1309.5626 [cond-mat.supr-con] (2013).

\bibitem{APL}
A. Gurevich, App. Phys. Lett. 88, 012511 (2006).

\end{thebibliography}

\end{document}